\documentclass[sigconf]{acmart}
\usepackage{balance}
\def\BibTeX{{\rm B\kern-.05em{\sc i\kern-.025em b}\kern-.08emT\kern-.1667em\lower.7ex\hbox{E}\kern-.125emX}}

\setcopyright{acmcopyright}

\usepackage[named]{algo}
\usepackage{graphicx}					
\usepackage{amsmath}
\usepackage{algorithm}
\usepackage{algorithmic}
\usepackage{ulem}
\usepackage{color}
\usepackage{verbatim}
\usepackage{multirow}
\usepackage{graphicx}
\usepackage{subfigure}
\usepackage{graphics}
\usepackage{balance}

\def\x{{\mathbf x}}

\def\R{\rm I\!R}
\def\N{\rm I\!N}
 
\providecommand{\abs}[1]{\lvert#1\rvert}

\definecolor{light}{gray}{.9}

\newcommand{\Comment}[1]{}

\newcommand{\offset}{\textsc{Offset}\xspace}

\newcommand{\argmin}{\operatornamewithlimits{argmin}}

\begin{document}
\fancyhead{}
\copyrightyear{2019}
\acmYear{2019}
\acmConference[CIKM '19]{The 28th ACM International Conference on Information and
Knowledge Management}{November 3--7, 2019}{Beijing, China}
\acmBooktitle{The 28th ACM International Conference on Information and Knowledge
Management (CIKM '19), November 3--7, 2019, Beijing, China}
\acmPrice{15.00}
\acmDOI{10.1145/3357384.3357801}
\acmISBN{978-1-4503-6976-3/19/11}

\title{Soft Frequency Capping for Improved Ad Click Prediction in Yahoo Gemini Native}

 \author{Michal Aharon, Yohay Kaplan, Rina~Levy, Oren~Somekh}
 \affiliation{Yahoo Research, Haifa, Israel}
 \email{{michala,yohay,rina.levy,orens}@verizonmedia.com}
 \author{Ayelet Blanc, Neetai Eshel, Avi Shahar, Assaf~Singer, Alex Zlotnik}
 \affiliation{Tech Yahoo, Tel Aviv, Israel}
  \email{{ablanc,neetai,avis,assafs,alexz}@verizonmedia.com}
  \renewcommand{\shortauthors}{Aharon, et al.}

\begin{abstract}
Yahoo's native advertising (also known as Gemini native) serves billions of ad impressions daily, reaching a yearly run-rate of many hundred of millions USD. Driving the Gemini native models that are used to predict both click probability (pCTR) and conversion probability (pCONV) is \offset\ -- a feature enhanced collaborative-filtering (CF) based event prediction algorithm. \offset is a one-pass algorithm that updates its model for every new batch of logged data using a stochastic gradient descent (SGD) based approach. Since \offset represents its users by their features (i.e., user-less model) due to sparsity issues, rule based hard frequency capping (HFC) is used to control the number of times a certain user views a certain ad. Moreover, related statistics reveal that user ad fatigue results in a dramatic drop in click through rate (CTR). Therefore, to improve click prediction accuracy, we propose a soft frequency capping (SFC) approach, where the frequency feature is incorporated into the \offset model as a user-ad feature and its weight vector is learned via logistic regression as part of \offset training. Online evaluation of the soft frequency capping algorithm via bucket testing showed a significant $7.3$\% revenue lift. Since then, the frequency feature enhanced model has been pushed to production serving all traffic, and is generating a hefty revenue lift for Yahoo Gemini native. We also report related statistics that reveal, among other things, that while users' gender does not affect ad fatigue, the latter seems to increase with users' age.   
\end{abstract}



\maketitle

\section{Introduction}\label{sec:Introduction}

\begin{figure}[!t]
\centering
\includegraphics[width=1.0\columnwidth]{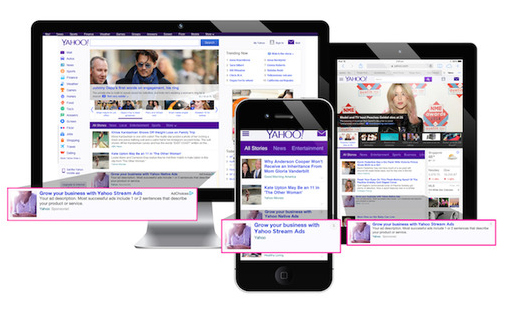}
\caption{Gemini native ads on different devices.}
\label{fig:Gemini native}
\end{figure}

Yahoo's native ad marketplace (also known as \textit{Gemini native} \footnote{https://gemini.yahoo.com/advertiser/home}) serves users with ads that are rendered to resemble the surrounding native content (see Figure \ref{fig:Gemini native} for examples). In contrast to the search-ads marketplace, users' intents are generally unknown. Launched five years ago and operating with a yearly run-rate of several hundred millions USD, Gemini native is one of Yahoo's main businesses. With more than two billion impressions daily, and an inventory of a few hundred thousand active ads, this marketplace performs real-time \textit{generalized second price} (GSP) auctions that take into account ad targeting, budget considerations, and frequency and recency rules, with SLA (or latency) of less than 80 ms more than 99\% of the time.

In order to rank native ads for an incoming user and their specific context according to the cost per click (CPC) price type, a score (or expected revenue) is calculated by multiplying the advertiser's bid and the predicted click probability (pCTR) for each ad. Although Gemini native handles other price types such as conversion (oCPC), in this work we focus on CPC price type.

The pCTR is calculated using models that are periodically updated by \offset\ - a feature enhanced collaborative-filtering (CF) based event-prediction algorithm \cite{aharon2013off}. \offset is a one-pass algorithm that updates its latent factor model for every new mini-batch of logged data using a \textit{stochastic gradient descent} (SGD) based learning approach. \offset is implemented on the grid using \textit{map-reduce} architecture \cite{dean2008mapreduce}, where every new mini-batch of logged data is preprocessed and parsed in parallel by many \textit{mappers} and the ongoing training of model instances with different hyper parameters sets is done in parallel by many \textit{reducers} to facilitate \offset adaptive online hyper-parameter tuning process \cite{aharon2017adaptive}.

\offset represents its users by their features (e.g., age, gender, geo, etc.), where each feature value (e.g., female, male, and unknown for the gender feature) is represented by a \textit{latent factor vector} (LFV). A user's LFV is derived from the user features' LFV by applying a non-linear function which allows for pairwise feature dependencies. Since \offset is a user-less model, the number of times a certain user views a certain ad (or frequency feature) cannot be captured by merely training the model over the logged impressions. Moreover, the frequency is neither a user- nor an ad- feature. Therefore, to prevent users from viewing the same ads over and over again, a rule based \textit{hard frequency capping} (HFC) is applied by the serving system during the ad ranking process. In general, ads that the user saw more than a predefined number of times during a predefined period are removed from the ranked list and are not allowed to participate in the auction.

Motivated by observations showing click-through rate (CTR) is decreasing with repeated ad views (see \cite{lee2014modelingImpressionDiscount}\cite{ma2016userFatigue}), in this work we consider a new approach to handle the frequency internally by the model treating it as a user-ad feature. According to this approach, referred to as \textit{soft frequency capping} (SFC), for each impression the frequency feature is calculated for the user-ad pair and used to train a frequency weight vector as part of \offset stochastic gradient descent (SGD). During serving, the appropriate weight is selected according to the frequency feature of the incoming impression and added to the \offset score. As we shall see, the frequency weight vector and therefore the resulting pCTR decrease with frequency, expressing the user fatigue of viewing the same ads repeatedly. Offline and online evaluations of the proposed approach reveal a staggering performance lift when comparing the SFC to the HFC. In particular, we measure a $7.3\%$ revenue lift for the online experiment serving real users, which translates into additional revenue of many millions of USD yearly. The SFC enhanced \offset model was pushed to production over a year ago and it has been serving all Gemini native traffic since. We also provide statistics gathered for the frequency feature, demonstrating the effect of the latter on the click tendency of different crowds. In general, user fatigue is observed in most settings, as the \textit{click through rate} (CTR) decreases with increasing frequency. Other more specific observations reveal that while male and female users experience similar ad fatigue patterns, fatigue increases drastically with age. In particular, we measure almost twice as much user fatigue among user of age group $50$-$60$ than that of age group $20$-$30$, after viewing a campaign five times.

The main contributions of this work are:
\begin{itemize}
\item Comprehensive statistics of the ad view frequency feature as it is manifested in the logged data of a web scale online advertising system. In addition, several interesting observations are made regarding the way different crowds are affected by the frequency feature.
\item The soft frequency capping approach (SFC) - a simple yet effective approach for incorporating the frequency feature into a user-less model such as \offset.
\item A thorough performance evaluation consisting of both offline and online (i.e., serving real Yahoo Gemini native users) evaluations, demonstrating the overwhelming superiority of the SFC approach over the previous rule based HFC approach. 
\end{itemize}

The rest of the paper is organized as follows. In Section~\ref{sec:Background}, we provide relevant background, and discuss related work in Section~\ref{sec:related work}. The frequency feature is described in Section \ref{sec: Frequency Feature}. Section \ref{sec: Frequency feature statistic and observations} is all about statistics and observations of the frequency feature. We set our goal in Section \ref{sec: Our Goal} and elaborate on our approach in Section \ref{sec:Our Approach}. Performance evaluation of our solution is presented in Section \ref{sec:Performance Evaluation}. We conclude and consider future work in Section~\ref{sec:Concluding Remarks and Future work}.

\section{Background}\label{sec:Background}
\subsection{Gemini Native}\label{sec: Gemini Native}


\begin{figure}[!t]
\centering
\includegraphics[width=1.0\columnwidth]{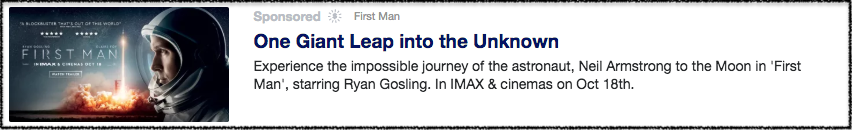}
\caption{A typical Gemini native ad from Yahoo home-page stream. The ad consists of a title, an image, a description, and a sponsorship notification.}
\label{fig:native ad}
\end{figure}


Gemini native is one of Yahoo's major businesses, reaching a yearly run-rate of a few hundred millions USD in revenue. Gemini native serves a daily average of more than two billion impressions world wide, with SLA (or latency) of less than $80$ ms for more than $99\%$ of the queries, and a native ad inventory of several hundred thousand active ads on average. Native ads resemble the surrounding page items, are considered less intrusive to the users, and provide a better user experience in general (see Figure \ref{fig:native ad} for a typical Gemini native ad on Yahoo home-page stream).

The online serving system is comprised of a massive Vespa \footnote{Vespa is Yahoo's elastic search engine solution.} deployment, augmented by ads, budget and model training pipelines. The Vespa index is updated continuously with ad and budget changes, and periodically (e.g., every $15$ minutes) with model updates. The Gemini native marketplace serves several ad price-types including CPC (cost-per-click), oCPC (optimizing for conversions), CPM (cost-per-thousand impressions), and also includes RTB (real-time biding) in its auctions. 

\subsection{The {OFFSET} Click-Prediction Algorithm}\label{sec:offset}
The algorithm driving Gemini native models is \offset: a feature enhanced collaborative-filtering (CF)-based ad click-prediction algorithm \cite{aharon2013off}. The \textit{predicted click-through-rate} (pCTR) of a given user $u$ and ad $a$ according to \offset is given by
\begin{equation}\label{eq: pCTR}
    \mathrm{pCTR}(u,a) = \sigma(s_{u,a})\in [0,1]\ ,
\end{equation}
where $\sigma(x)=\left(1+e^{-x}\right)^{-1}$ is the sigmoid function, $s_{u,a}=b+\nu_{u}^T\nu_{a}$ ,
and $\nu_{u},\nu_{a}\in \R^D$ denote the user and ad latent factor vectors respectively, and $b\in \R$ denotes the model bias. The product $\nu_{u}^T\nu_{a}$ denotes the tendency score of user $u$ towards ad $a$, where a higher score translates into a higher pCTR. Note that $\Theta=\{\nu_{u},\nu_{a},b\}$ are model parameters learned from logged data, as explained below.

Both ad and user vectors are constructed using their features, which enable dealing with data sparsity issues (ad CTR is quite low in general). For ads, we use a simple summation between the vectors of their features (e.g., unique creative id, campaign id, advertiser id, ad categories, etc.), all in dimension $D$. The combination between the different user feature vectors is a bit more complex in order to allow non-linear dependencies between feature pairs. 


The user vectors are 
constructed using their $K$-feature latent vectors $v_k\in \R^d$ (e.g., age, gender, geo, etc.). In particular, $o$ entries are devoted for each pair of user
feature types, and $s$ entries are devoted for each feature type vector alone.
The dimension of a single feature value vector is therefore $d=(K-1)\cdot o + s$,
whereas the dimension of the combined user vector is $D={K\choose 2} \cdot o + K\cdot s$.
The advantage over the standard CF approach is that the model includes only $O(K)$ feature LFVs (one for each feature value, e.g., 3 for gender - male, female and unknown) instead of hundreds of millions of unique user LFVs.

To learn the model parameters $\Theta$, \offset minimizes the logistic loss (LogLoss) of the training data set $\mathcal{T}$ (i.e., past impressions and clicks) using one-pass \textit{stochastic gradient descent} (SGD). The cost function is as follows:
\begin{equation*}
\argmin_{\Theta}\!\!\!\! \sum_{(u,a,y,t)\in \mathcal{T}} \mathcal{L}(u,a,y,t)\ ,
\end{equation*}
where
\begin{multline*}
\mathcal{L}(u,a,y,t)=-(1-y)\log\left(1-pCTR(u,a)\right)\\-y \log pCTR(u,a)+\frac{\lambda}{2}\sum_{\theta\in\Theta}\theta^2\ ,
\end{multline*}
$y \in \{0,1\}$ is the click indicator for the event involving user $u$ and ad $a$ at time $t$, and $\lambda$ is the $L2$ regularization parameter. For each training event $(u,a,y,t)$, \offset updates its relevant model parameters by the SGD step
\[
\theta\gets\theta-\eta(\theta)\triangledown_\theta\mathcal{L}(u,a,y,t)\ ,
\]
where $\triangledown_\theta\mathcal{L}(u,a,y,t)$ is the gradient of the objective function w.r.t $\theta$. In addition, the parameter-dependent step size is given by
\[
\eta(\theta)=\eta_0\frac{1}{\alpha+\left(\sum_{(u,a,y,t)\in \mathcal{T}'}\abs{\triangledown_\theta\mathcal{L}(u,a,y,t)}\right)^\beta}\ ,
\]
 where $\eta_0$ is the SGD initial step-size, $\alpha,\ \beta\in\R^+$ are the parameters of our variant of the adaptive gradient (AdaGrad) algorithm \cite{duchi2011adaptive}, and $\mathcal{T}'$ is the set of training impressions seen so far.

The \offset algorithm uses an online approach where it continuously updates its model parameters with each batch of new training events (e.g., every 15 minutes for the click model). A more elaborate description, including details on using AdaGrad \cite{duchi2011adaptive}, multi-value features, and regularization can be found in \cite{aharon2013off}. 

The \offset algorithm includes an adaptive online hyper-parameter tuning mechanism \cite{aharon2017adaptive}. This mechanism takes advantage of the parallel map-reduce architecture and strives to tune \offset hyper-parameters (e.g., SGD initial step size and AdaGrad parameters) to match the varying marketplace conditions (changed by temporal and trend effects).


\subsection{Serving}\label{sec: Serving}

When a user arrives at a Yahoo \textit{owned and operated} (O\&O) or Syndication \footnote{Where Yahoo presents its ads on a third party site and shares the revenue with the site owner.} site, and a Gemini native slot should be populated by ads, an auction takes place.
Initially, Serving generates a list of eligible active ads for the user as well as each ad's score.
Roughly speaking, an ad's eligibility to a certain user in a certain context is determined by targeting, and HFC rules. 

\paragraph{Hard frequency capping}
While targeting, which is outside the scope of this work, relates to user characterization (e.g., age, gender, geo, etc.) specified by the advertiser to only approach certain crowds, HFC limits the number of times a certain user sees a certain ad. In particular, Gemini native serving uses the following simple HFC rules to prevent a user from repeatedly seeing the same ads:
\begin{itemize}
\item A user cannot view ads of a certain campaign more than five times a week.
\item A user cannot view a certain ad more than twice a day.
\end{itemize}
Note that these are default values and the advertiser may alter the numbers to suite her needs via the Gemini platform.
As mentioned earlier, this work is all about replacing HFC with and improved SFC solution. 

\paragraph{Auction}
The score is a measure that attempts to rank the ads according to their potential revenue with respect to the arriving user and her context (e.g., day, hour, site, device type, etc.). In general, an ad's score is defined as
$
\mathrm{rankingScore}(u,a) = \mathrm{bid}(a) \cdot \mathrm{pCTR}(u,a)
$
where pCTR (predicted click through rate) is provided by an \offset model (see Eq. \eqref{eq: pCTR}), and $\mathrm{bid}(a)$ is the amount of money the advertiser is wiling to pay for a click on ad $a$.

To encourage advertiser truthfulness, the cost incurred by
the winner of the auction is according to \textit{generalized second price} (GSP) \cite{edelman2007internet}, which is defined as
\begin{equation}
\label{eq:gsp}
\mathrm{gsp} = \frac{\mathrm{rankingScore}_2}{\mathrm{rankingScore}_1} \cdot \mathrm{bid}_1 = \frac{\mathrm{pCTR}_2}{\mathrm{pCTR}_1} \cdot \mathrm{bid}_2 \ ,\nonumber
\end{equation}
where indices $1$ and $2$ correspond to the winner of the auction and the runner up, respectively. By definition $\mathrm{gsp}\le \mathrm{bid}_1$, so the winner will pay no more than its bid. 

\section{Related work and practice}\label{sec:related work}

Recommendation technologies are essential for CTR prediction. \textit{Collaborative filtering} (CF) in general and specifically \textit{matrix factorization} (MF) based approaches are leading recommendation technologies, according to which entities are represented by latent vectors and learned by users' feedback (such as ratings, clicks and purchases) \cite{koren2009matrix}. MF-CF based models are used successfully for many recommendation tasks such as movie recommendation \cite{bell2007lessons}, music recommendation \cite{aizenberg2012build}, ad matching \cite{cs:off_set}, and much more. CF is evolving constantly, where recently it was combined with \textit{deep learning} (DL) for embedding entities into the model \cite{he2017neural}.

There are few published works describing models driving web scale advertising platforms. In \cite{mcmahan2013ad} lessons learned from experimenting with a large scale logistic regression model (LSLR) used for CTR prediction by Google advertising system are reported. These include improvements in traditional supervised learning based on a \textit{follow the regularized leader} (FTRL)-like online learning \cite{mcmahan2011follow}, and the use of per-coordinate learning rates.  A model that combines decision trees with logistic regression is used to drive Facebook CTR prediction and is reported on in \cite{he2014practical}. The authors conclude that the most important thing is to have the right features. Specifically, those capturing historical information about the user or ad dominate other types of features. Placing ads in a tweet stream is  considered in \cite{li2015click}, where pairwise ranking is used to train a LSLR model for CTR prediction. In a way this task resembles the problem of native ad CTR prediction since user intention is not clear here as well.

Yahoo has also shared its native ad click prediction algorithm with the community where an earlier version of \offset was presented \cite{cs:off_set}. A mature version of \offset was presented in \cite{aharon2017adaptive}, where the focus was on the adaptive online hyper-parameter tunning approach of it, taking advantage of its parallel system architecture. Unlike the three commercial CTR prediction LSLR models mentioned earlier, \offset is a feature enhanced CF based model. \Comment{The fact that two recent Kaggle CTR prediction challenges \footnote{https://www.kaggle.com/c/criteo-display-ad-challenge\\ https://www.kaggle.com/c/avazu-ctr-prediction} were won by \textit{field aware factorization machine} (FFM) -like models (see \cite{rendle2010pairwise} \cite{juan2016field}\cite{jahrer2012ensemble}) may suggest that \offset, which can be seen as a special case of FFM, is a worthy candidate for the task.}


Frequency capping has previously been studied as user fatigue in recommendation systems which  happens when users are repeatedly shown the same items. In \cite{agarwal2009spatio} the fatigue issue is considered for content recommendation. The authors noticed the CTR drop with repeated views and adjust their models for user fatigue through an exponential tilt to the first-view CTR (probability of click on first article exposure) that is based only on user-specific repeat-exposure features. Another work dealing with user fatigue in news recommendation is \cite{ma2016userFatigue}. After analyzing the user fatigue on Microsoft Bing Now news recommendation service from different perspectives such as demographics (i.e., age, and gender), the authors proposed features that are correlated with fatigue related CTR variations. Then, they demonstrated the benefit of using these features for providing improved recommendations. In \cite{lee2014modelingImpressionDiscount} the authors study the user fatigue impact on LinkedIn ``People You May Know'' and skills endorsement recommendations. Then they use the number of views and time since last view to calculate an impression discount factor that may be used as an external plug-in to an existing recommender system (as opposed to our in-model approach). A frequency Capping concept in online advertising, which limits user exposure to an ad due to the advertiser budget constraints is considered in \cite{buchbinder2014frequency}. The problem is formulated as an optimization problem which maximizes an advertiser's value by fixing a user's frequency cap and imposing some other constraints.

Our work is different from the above in several aspects. It deals with user ad fatigue which is quite different than user fatigue related to content recommendation (in general, ads are less tolerated by users than regular content). Therefore, the presented statistics, observations, and treatment are novel. For example, while \cite{ma2016userFatigue} reports that users' age has little effect on content fatigue, we report the opposite showing that age is a major factor in user ad fatigue. In our approach the frequency feature is included in the model and is learned as the rest of its parameters, while in \cite{lee2014modelingImpressionDiscount} an external plug-in solution is considered. Moreover, while most papers use datasets to test their models, this paper reports the performance of the frequency enhanced model in an online setting, serving ads to real users.  

\section{Frequency Feature}\label{sec: Frequency Feature}
The logged activity \footnote{In compliance with European and US privacy laws which are beyond the scope of this work.} of Yahoo's users in its O\&O and syndication properties also includes native ad impressions from which we can extract and calculate the frequency, i.e., the number of times a specific user has seen a certain ad during a predefined period of time. We can calculate the frequency for each ad feature (e.g., creative, campaign or advertiser). Therefore, after setting the ad feature $A_f$, and time period $T_f$, we can provide for each user $u$ and each ad $a$ the frequency feature $f_{u,a}(A_f,T_f)$ (or in short $f_{u,a}$). It is noted that by definition the frequency feature is a non-negative integer $f_{u,a}\in \N^+$.
\paragraph{Example}
Assume that user $u$ has seen three ads $a_1$, $a_2$, and $a_3$, each ad $a_i$ has the ad features: advertiser $Ad_i$, campaign $Ca_i$ and creative $Cr_i$ 
Moreover, assume that it's Saturday night just after midnight and user $u$'s Gemini native daily activity log during the last week is given in Table \ref{table:Activity log} (where time moves left to right).
\begin{table}  
\begin{center}
    \begin{tabular}{| c || c | c | c |c | c | c | c |} 
    \hline
    Ad & Su & M &  Tu & W & Th & F & Sa\\
    \hline \hline
    $a_1$ & $0$ & $0$ & $0$ & $1$ & $0$ & $0$ &  $2$\\
    \hline
    $a_2$ & $1$ & $1$ & $0$ & $0$ & $0$ & $2$ &  $1$\\
    \hline
    $a_3$ & $0$ & $0$ & $1$ & $1$ & $2$ & $0$ &  $1$\\
    \hline
    \end{tabular} 
    \end{center}
    \caption{Activity log}\label{table:Activity log}
\end{table}
Following are a few values of the frequency feature in different settings.
\[
\begin{aligned}
f_{u,a_1}(\mathrm{camp.},\mathrm{last\ day})=2\ &;\ f_{u,a_1}(\mathrm{adver.},\mathrm{last\ day})=3\\
f_{u,a_2}(\mathrm{camp.},\mathrm{last\ week})=5\ &;\ f_{u,a_2}(\mathrm{adver.},\mathrm{last\ day})=3\\
f_{u,a_3}(\mathrm{adver.},\mathrm{last\ 4\ days})=4\ &;\ f_{u,a_3}(\mathrm{adver.},\mathrm{last\ week})=5\ .
\end{aligned}
\]


\section{Statistics and observations}\label{sec: Frequency feature statistic and observations}

In this section we present some statistics and observations regarding the frequency feature. Most importantly, we show that the frequency feature is significant and has strong impact on the CTR. The statistics were aggregated during $30$ days earlier this year, over a portion of Yahoo Gemini traffic. It includes many billions of impressions and clicks. We note that the data used here was collected when the SFC approach was already included in \offset, serving all traffic.

\paragraph{Global view}

\begin{figure}[!htb]
\centering
\includegraphics[width=1.0\columnwidth]{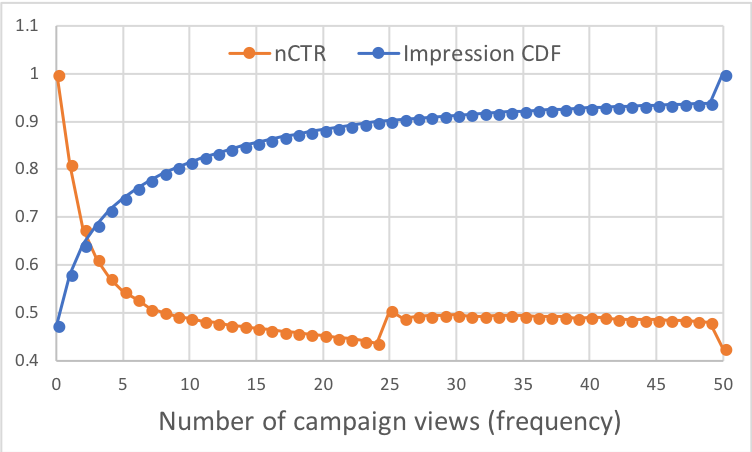}
\caption{Normalized CTR and impression CDF vs. number of campaign views (frequency).}
\label{fig: global ctr cdf}
\end{figure}
In Figure \ref{fig: global ctr cdf}, the average normalized CTR \footnote{Due to commercial confidentiality matters, we cannot share absolute numbers of traffic size, and performance.}, and the number of impressions \textit{cumulative density function} (CDF), are plotted as functions of the number of views $v$ (or frequency) of a certain campaign by a certain user. Note, that $v=0$ means that in these impressions the  users haven't seen the presented campaigns before. The normalized CTR is calculated by dividing the average CTR, measured for $v$ views, by the average CTR measured for no previous views $v=0$
\[
\mathrm{CTR}_n(v)=\frac{\mathrm{CTR}(v)}{\mathrm{CTR}(0)}\quad ; \quad v=0,1,\ldots,50\ .
\]
It is noted that in both curves, the last point includes all measurements aggregated for $v\ge 50$.

Examining the figure, several observations can be made. Putting aside the anomaly at $v=25$ \footnote{We will provide an explanation for this anomaly (repeated in most curves presented in this section) after we describe our approach of incorporating the frequency feature into \offset, and the way it affects the reported statistics.}, the CTR decreases monotonically with the number of views (or frequency). Specifically, the average CTR drops by $20\%$ after only a single past view, and almost by $50\%$ after $7$ views. This is a clear evidence of users' rapid fatigue, seeing the same campaign ads over and over again. However, the CTR descent rate decreases with the number of views, and the number of impressions decreases with the frequency (ignoring the last point which includes all impressions with $v\ge 50$). 
In particular, $47\%$ of the impressions are of never-seen-before campaigns (v=0), $10\%$ are for campaigns that have been seen-once-before ($v=1$), and only $6\%$ of the impressions are of campaigns seen-twice-before ($v=2$).

\paragraph{Gender view}

\begin{figure}[!htb]
\centering
\includegraphics[width=1.0\columnwidth]{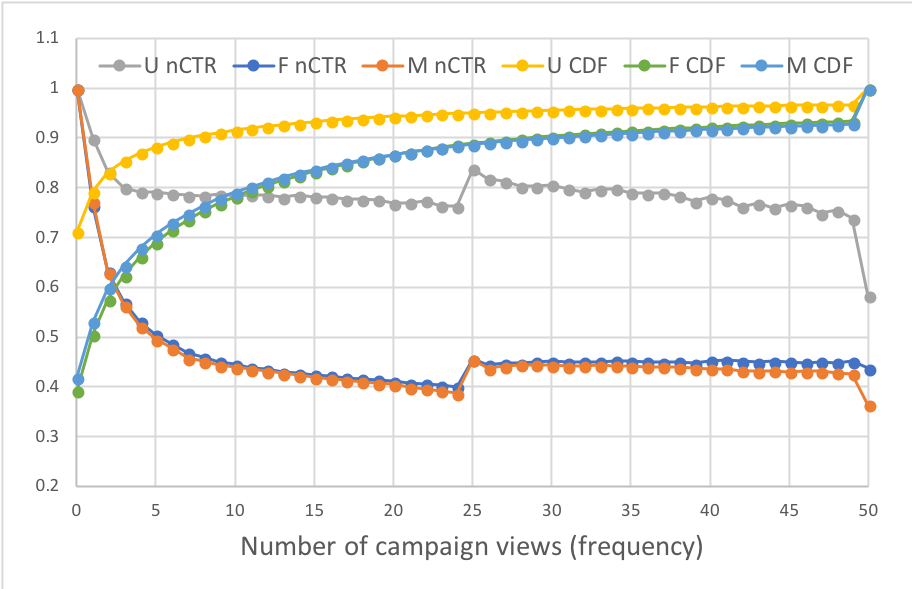}
\caption{Normalized CTR and impression CDF vs. number of campaign views (frequency) per gender.}
\label{fig: gender ctr}
\end{figure}

\begin{table}  
\begin{center}
    \begin{tabular}{| c | c | c | c |} 
    \hline
    Gender & Female & Male & Unknown\\
    \hline
    Traffic & $31.4\%$ & $46.8\%$ & $21.8\%$ \\
    \hline
    \end{tabular} 
    \end{center}
    \caption{Gemini native traffic share of genders.}\label{table: gender}
\end{table}

In Figure \ref{fig: gender ctr} the normalized CTR and impressions CDF are plotted as functions of the frequency for females, males, and users with unknown gender (identified, if at all, via their HTTP cookies). The Gemini native traffic share (in terms of ad impressions) of each gender is presented in Table \ref{table: gender}. Surprisingly, there are many more declared male users then female users.
Gender uncertainty is due to registered users that do not declare their gender, and mostly due to unregistered users' activity. The impressions CDF curves provide support for the latter, revealing that $70\%$ of the unknown users impressions are of never-seen-before ($v=0$) campaigns, while both male and female have only $~40\%$ of such impressions. 

Examining the figure we observe that frequency has almost the same affect over both male and female users, which demonstrate almost identical fatigue patterns. However, users with an unknown gender behave quite differently, demonstrating much higher tolerance to ad repeated views. A plausible explanation for such behavior is that these users, which are likely to be unregistered users, arrive to Yahoo properties from external search or social media sites and have a different experience than the registered users (e.g., visiting mostly certain Yahoo properties with better ad experience).
 
\paragraph{Age group view}
 
\begin{figure}[!htb]
\centering
\includegraphics[width=1.0\columnwidth]{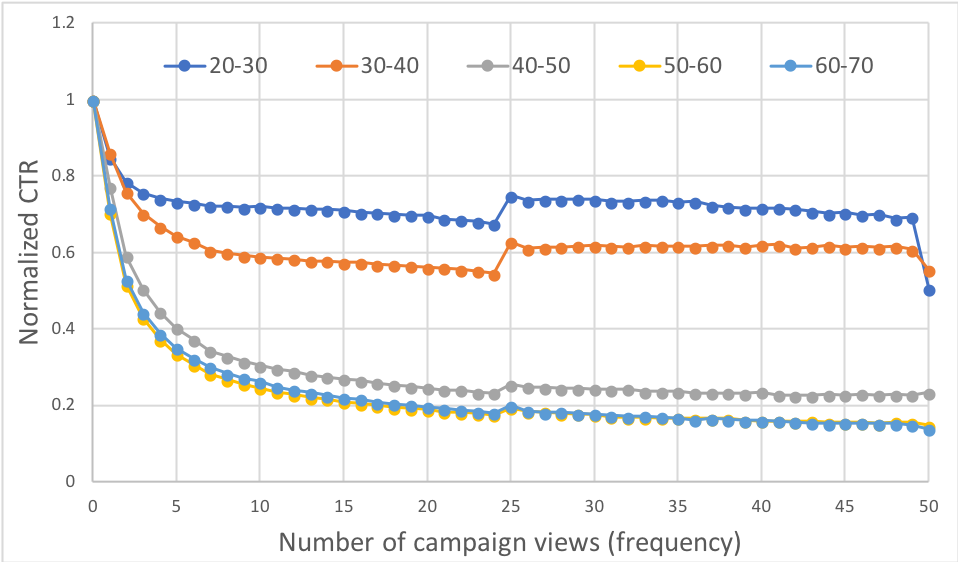}
\caption{Normalized CTR vs. number of campaign views (frequency) for several user age groups.}
\label{fig: age ctr}
\end{figure}

\begin{table}  
\begin{center}
    \begin{tabular}{| c | c | c | c | c | c |} 
    \hline
    Age & $20-30$ & $30-40$ & $40-50$ & $50-60$ & $60-70$\\
    \hline
    Traffic & $19.5\%$ & $35.8\%$ & $19.0\%$ & $10.9\%$ & $6.4\%$ \\
    \hline
    \end{tabular} 
    \end{center}
    \caption{Gemini native traffic share of several age groups}\label{table: age}
\end{table}

In Figure \ref{fig: age ctr} the normalized CTR is plotted as function of the campaign frequency for several age groups. The Gemini native traffic share of these age groups are reported in Table \ref{table: age}. Examining the figure it is observed that the normalized CTR decreases with campaign view frequency for all age groups. Moreover, it decreases much faster with user age. In particular, \textit{user fatigue} (measured here by the drop in normalized CTR) after seeing a campaign 5 times, is more than twice for user age group $50$-$60$ when compared to that of user age group $20$-$30$. The observation that user fatigue increases with age is a bit counterintuitive and surprising. A possible explanation for this observation, is that younger users are more used to having ads as part of their Internet browsing experience since childhood. It is also noted that while user fatigue increase with age, its increase rate reduces eventually, where age groups $50$-$60$ and $60$-$70$ exhibit almost identical fatigue pattern.       

\paragraph{Yahoo vertical view}

\begin{figure}[!htb]
\centering
\includegraphics[width=1.0\columnwidth]{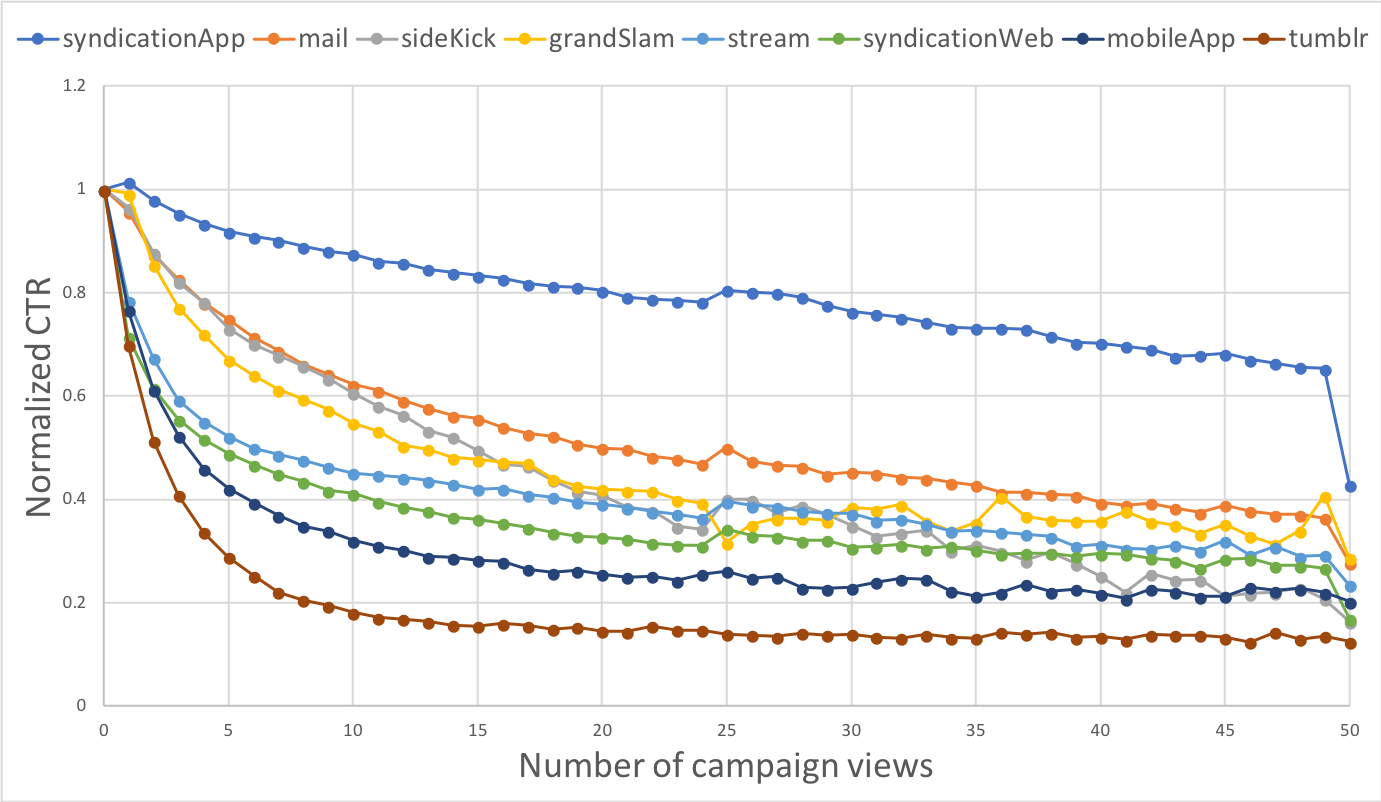}
\caption{Normalized CTR vs. number campaign views for several Yahoo verticals.}
\label{fig: vertical ctr}
\end{figure}

\begin{table}  
\begin{center}
    \begin{tabular}{| c | c | c |} 
    \hline
    Yahoo vertical & Description & Traffic \\
    \hline \hline
    GrandSlam & Yahoo articles and news & $6.0\%$ \\
    \hline
    Mail & Yahoo mail & $32.3\%$ \\
    \hline
    MobileApp & Yahoo mobile apps & $3.5\%$ \\
    \hline
    SideKick & Article recommendation property & $3.6\%$ \\
    \hline
    Stream & Yahoo home page & $16.6\%$ \\
    \hline
    SyndicationApp & Third party mobile apps & $21.5\%$ \\
    \hline
    SyndicationWeb & Third party web sites & $12.0\%$ \\
    \hline
    Tumblr & Tumblr & $2.3\%$ \\
    \hline
    \end{tabular} 
    \end{center}
    \caption{Gemini native traffic share of several Yahoo verticals}\label{table: vertical}
\end{table}

In Figure \ref{fig: vertical ctr} the normalized CTR is plotted as function of the campaign frequency for several of Yahoo verticals. A vertical is an arbitrary collection of page sections that have something in common although they may be served on different devices and can be quite different in appearance. While some of them are self explanatory by their name (i.e., mail and tumblr) others are more opaque and require intimate knowledge of Yahoo properties. The verticals' definitions along with their Gemini native traffic share are elaborated in Table \ref{table: vertical}.

Since verticals are quite different from one another, it is hard to draw conclusions from their relative users' figures, beside stating the obvious that tumblr users demonstrates the strongest ad fatigue while syndicationApp users show the weakest ad fatigue. Having a weak fatigue may also indicate the level of ad ``blindness'' users demonstrate in certain verticals such as Yahoo mail (the second weakest user fatigue vertical), where native ads are placed at the top of the mail timeline, and get little attention by mail users who are focused on their mail correspondence. It is also worth mentioning that two verticals (syndicationApp, and grandSlam) demonstrate somewhat different behavior than other verticals, where their nCTR for seen-once ($v=1$) is higher or comparable than their never-seen ($v=0$) nCTR. This may be explained since many of these verticals' users are non-registered users making a single visit to these properties (arriving from external search or social media sites) and act differently than registered users with multiple visits.      

\section{Our Goal}\label{sec: Our Goal}

The goal of this project, is to adopt a \textit{soft frequency capping} (SFC) approach by incorporate the frequency feature into the \offset model. The proposed solution should be optimized to provide ``best'' offline and online performance (performance metrics will be specified in Section \ref{sec:Performance Evaluation}), and outperform the legacy \textit{hard frequency capping} (HFC) solution.

\section{Soft Frequency Capping}\label{sec:Our Approach}
\paragraph{Overview}
The frequency feature (see Section \ref{sec: Frequency Feature}) is simply the number of times a specific user has seen a certain predefined ad feature $A_f$ (creative, campaign, or advertiser) in a predefined time period $T_f$ (e.g., last day, last week, or last month). In this section we present our approach to integrate this feature into the \offset model. 

In general, we consider the frequency feature as a user-ad feature, where we learn a frequency weight vector(s) according to a predefined weights category parameter $W_c$ which determines if we have a single global vector or a separate vector per campaign or per advertiser. 

In particular, for each incoming train of events $\{(u,a,y,t)\}$, the feature value $f_{a,u}(A_f,T_f)$ is binned, multiplied by the corresponding entry of the appropriate frequency weight vector, and added to the \offset score. The frequency weight vectors are learned as part of the SGD described in Section \ref{sec:offset} using the user and ad features, and the label $y$ (click or skip). In serving time the frequency weight vectors are used as part of the \offset model to calculate the pCTR to be used during Yahoo Gemini native auctions.  


\paragraph{Formal description}
A formal description of our SFC approach is elaborated in Algorithm 1.
\begin{algorithm}[!htb]
\caption{\offset soft frequency capping (SFC)} \label{algo:frequency modeling}
\textbf{Input:}\\
$A_f$ - ad feature (creative/campaign/advertiser)\\
$T_f$ - history window size (day/week/month)\\
$W_c$ - weights category (global/campaign/advertiser)\\
$B$ - binning operator\\
\textbf{Output (updated after each event):}\\
$\{{\bf w}\}$ - weight vectors\\
$\sigma(s'_{u,a})$ - \offset pCTR\\
\begin{algorithmic}[1]
\STATE initialize all weight vectors $\{\bf{w}\}$ entries with zeros 
\FOR{each event $(u,a,y,t)\in \mathcal{T}$}
\STATE calculate $f_{u,a}$ according to $A_f$ and $T_f$
\STATE get the weight vector $\bf{w}$ according to $W_c$ and $A_f$
\STATE calculate the bin index $i=B(f_{u,a})$
\STATE calculate the new \offset score
\STATE \hspace{0.4cm} $s'_{u,a}=s_{u,a}+{\bf w}_i$
\STATE calculate the gradient of $\mathcal{L}(u,a,y,t)$ w.r.t. ${\bf w}_i$  
\STATE \hspace{0.4cm} $\triangledown_{{\bf w}_i}\mathcal{L}(u,a,y,t)=\left(y-\sigma(s'_{u,a})\right)+\lambda {\bf w}_i$
\STATE update ${\bf w}_i$
\STATE \hspace{0.4cm} ${\bf w}_i \gets {\bf w}_i - \eta({\bf w}_i)\triangledown_{{\bf w}_i}\mathcal{L}(u,a,y,t)$
\ENDFOR
\end{algorithmic}
\end{algorithm} 

\paragraph{Why binning?}
As an alternative to the binning based approach, we could have used a linear regression for the additive frequency weight
\[
s'_{u,a}=s_{u,a}+c_a\cdot g(f_{u,a})
\]
where $c_a$ is a weight that can be learned globally, per campaign, or per advertiser, and $g(\cdot)$ is an arbitrary function. The advantage of having a weight vector (with a weight entry per bin) is that we do not assume that a certain dependency (i.e., $g(\cdot)$) provides the best performance, and we let the model ``decide'' what is the best fit. In our case there are no drawbacks either, since the frequency feature can have only non-negative integer values and quantization errors can be totally avoided. 

\paragraph{Expected impact}
It would perhaps be intuitive to think that the impact of such an approach would be confined to the scores given to the repeated (second and onwards) views of an ad by  the same users. However, theoretical considerations (and the eventual results) show that some of the impact is actually inflicted on the first views of an ad.

When using HFC, the scores of a predictive model that ignores frequency must tend towards an average of the CTR on first and repeated impressions. Since repeated impressions have lower CTR, these scores are lower than the CTR of the first views. Adding SFC allows the pCTR to be higher on the first view and decrease on later views due to the SFC weights \footnote{We show in Section \ref{fig: online lifts} that the SFC weights are decreasing with the number of views.}. Hence, the scores of ads that were receiving many multiple views are no longer deflated by those views, and that the click predictions of their first views are now more accurate, as well as those predictions of their later views.

\section{Performance Evaluation}\label{sec:Performance Evaluation}

In this section we report the offline and online performance of the SFC enhanced \offset model. For both cases we describe the setting, define the performance metrics, and present the results.

It is noted that since propriety logged data is used for evaluating our model, it is obvious that reproducing the results by others is impossible. This caveat is common in papers describing commercial systems and we hope it doesn't undermines the overall contribution of this work.  

\subsection{Offline Evaluation}\label{sec: Offline Evaluation}

\begin{figure}[!htb]
\centering
\includegraphics[width=1.0\columnwidth]{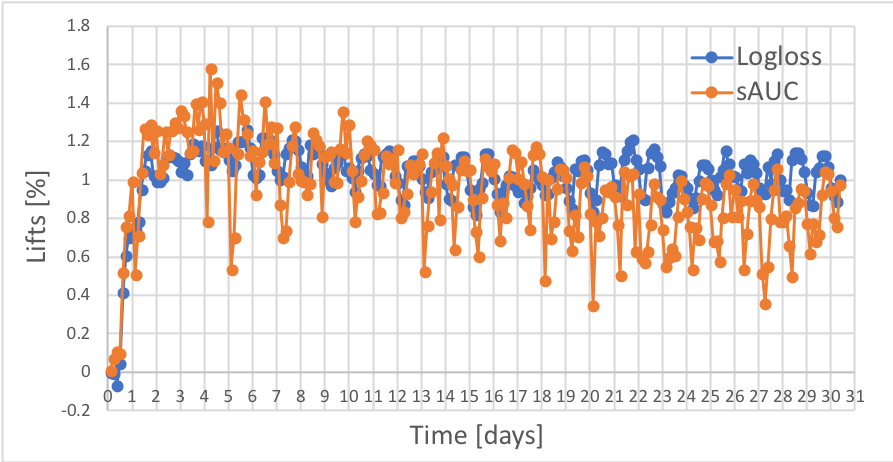}
\caption{Offline LogLoss and stratified AUC lifts of soft frequency capping vs. time in [days].}
\label{fig: offline lifts}
\end{figure}

\begin{figure}[!htb]
\centering
\includegraphics[width=1.0\columnwidth]{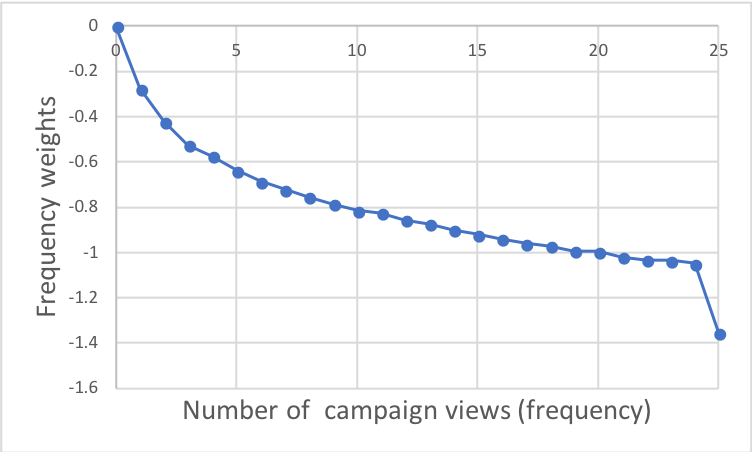}
\caption{Global campaign frequency weight vector.}
\label{fig: weight vector}
\end{figure}

\paragraph{Setup}
To evaluate offline performance we train two \offset models, one with SFC as described in Section \ref{sec:Our Approach} and the other with no frequency capping, serving as a baseline. We run both models from ``scratch'' where all model parameters are randomly initialized, over one month of Gemini native logged data, which includes many billions of impressions.

Due to technical limitations involving the serving system (and are beyond the scope of this work) we use the following binning vector with $26$ bins
\[
B_{26}=[0:1),[1:2),\ldots,[25:\infty)\ .
\]
Moreover, we have tested many combinations of the SFC algorithm parameters (not presented here) and found that the best setup is to use campaign as ad feature ($A_f=$ campaign), aggregate views over the last week ($T_f=$ week), and use a single global weight vector ($W_c=$ global) which eliminates any sparsity issues. We use both sAUC and LogLoss metrics (defined next), to measure offline performance, where each impression is used for training the system before being applied to the performance metrics. \offset hyper-parameters, such as SGD step size and regularization coefficient, are determined automatically by the adaptive online tuning mechanism included in \offset (see \cite{aharon2017adaptive}).

\paragraph{Performance metrics}\label{sec:Performance metrics}

\begin{description}
\item[Area-under ROC curve (AUC)] The AUC specifies the
probability that, given two random events (one positive and one negative, e.g., click and skip), their predicted pairwise
ranking is correct \cite{fawcett2006introduction}.
\item[Stratified AUC (sAUC)] The weighted average
(by number of positive event, e.g., number of clicks) of the AUC of each Yahoo section. This metric is used since different Yahoo sections have different prior click bias and therefore even using the section feature alone turns out as sufficient for achieving high AUC values.

\item[Logistic loss (LogLoss)]
\[
\sum_{(u,a,y,t)\in \mathcal{T}} \! \! -y \log
pCTR(u,a)-(1-y)\log\left(1-pCTR(u,a)\right),
\]
where ${\mathcal{T}}$ is a training set and $y \in \{0,1\}$ is the positive event indicator (e.g., click or skip). We note that the LogLoss metric is used to optimize \offset model parameters and its algorithm hyper-parameters.
\end{description}

\paragraph{Results}
The LogLoss and sAUC lifts \footnote{Since lower-is-better for LogLoss and higher-is-better for sAUC, the lifts in percentage are given by $(1-\mathrm{LogLoss_{SFC}}/\mathrm{LogLoss_{baseline}})\cdot 100\ $ and $\ (\mathrm{sAUC_{SFC}}/\mathrm{sAUC_{baseline}}-1)\cdot 100$, respectively.} are plotted vs. time in Figure \ref{fig: offline lifts} for an \offset model trained with binning vector $B_{26}$ and the best SFC algorithm parameters $A_f=$ campaign, $T_f=$ week, and $W_c=$ global, where each point represents $3$ hours worth of data. Examining the figure, the superiority of the SFC model over the baseline is evident and statistically sound with all reported lifts are positive. In particular, we measure an average $1.02\%$ LogLoss lift and $0.83\%$ sAUC lift over the last week of training. We note that achieving such high accuracy improvements for a mature algorithm such as \offset, which is continuously being optimized over several years now, is quite unexpected and impressive.

To complete this part we present the resulting global campaign frequency weight vector learned by the model in Figure. \ref{fig: weight vector}. As expected from the nCTR tendency observed in Section \ref{sec: Frequency feature statistic and observations}, the weights decrease monotonically with the frequency where the last point covering all frequencies larger than $v=25$ drops way below the extrapolated curve. The latter may cause pCTR inaccuracies for events occurring in this region, e.g., under-prediction for $f_{u,a}=25$ and over-prediction for $f_{u,a}\gg 25$. Since we have less events with higher frequencies (see Figure \ref{fig: global ctr cdf}) it is expected that the overall average effect would be of under-prediction.    

\paragraph{Statistics anomaly explained}
At this point the ground is set to explain the anomaly observed in most nCTR curves presented in Section \ref{sec: Frequency feature statistic and observations}. The anomaly consists of a small ``jump'' in the nCTR, which ``breaks'' the monotone nCTR descent with frequency, that occur between $v=24$ and $v=25$. As mentioned earlier, the statistics were collected when the SFC was already integrated into \offset using the binning vector $B_{26}$ with $[25:\infty)$ as last bin. As mentioned in the previous paragraph this may cause an overall under-prediction effect of events falling in this region. Since the statistics are collected only for auction winning events, we get that in-spite of the under-prediction for frequencies $v\ge 25$ these ads won the auction and when we record their true nCTR it is higher than expected and therefore we get this ``jump'' in nCTR.      

\subsection{Online Evaluation}\label{sec:Offline Evaluation}

\begin{figure}[!htb]
\centering
\includegraphics[width=1.0\columnwidth]{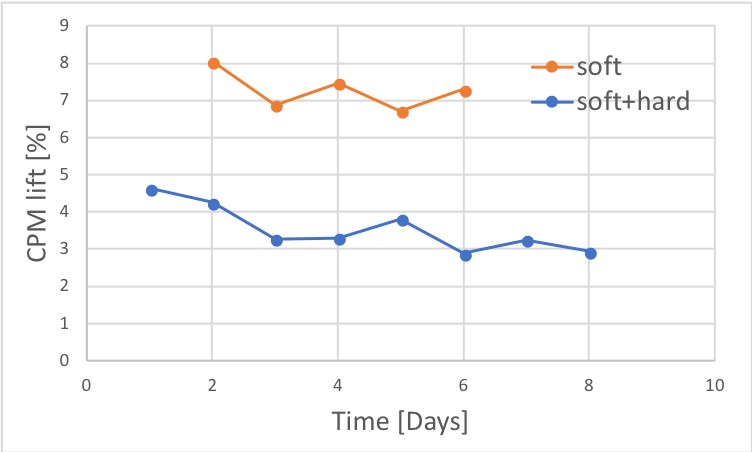}
\caption{Online CPM (cost per thousand impressions) lifts of hybrid frequency capping (hard + soft) and soft frequency capping vs. time in [days].}
\label{fig: online lifts}
\end{figure}

\paragraph{Setup}
To evaluate the online performance that determines whether the frequency feature enters the production model, we launched several online buckets serving a portion of Yahoo Gemini traffic and measured the revenue lifts in terms of the \textit{average cost per thousand impressions} (CPM) with respect to the production model. Recall that the production model at that time, which served as baseline, did not include the SFC, and rule based HFC was being imposed by the serving system (see Section \ref{sec: Serving}). For the SFC enhanced model we used the same parameters as the production model and added the frequency feature using $B_{26}$ and the best parameters set, i.e., $A_f=$ campaign, $T_f=$ week, and $W_c=$ global. Note that we were using ``mature'' SFC models that were trained from ``scratch'' on many weeks worth of data before allowed to serve real users.

Being cautious, we started by launching a bucket with a SFC model, serving $1\%$ of Yahoo Gemini native traffic, but kept the HFC rules. We refer to the combination of SFC model with HFC rules (imposed by the serving system) as the \textit{hybrid bucket}. After making sure the hybrid bucket ``behaves'' properly we launched a second bucket with SFC model and aborted the serving HFC rules from its portion of the traffic. Then, we gradually increased the traffic portions of both buckets. The results reported next were recorded when the hybrid bucket was serving $50\%$ of traffic and the SFC bucket was serving $5\%$ of traffic.  



\paragraph{Results}
The daily CPM lifts of both hybrid and SFC buckets when compared to the production bucket (operating with HFC only) are presented in Figure \ref{fig: online lifts} over several days. The figure reveals the staggering average CPM lifts of $3.5\%$, and $7.3\%$ measured for the hybrid and SFC buckets, respectively. 
Also notable is the impressive average CPM $3.6\%$ lift that occurs once we abort the HFC rules imposed by the serving system and remain with the SFC enhanced \offset model only. Such improvements translate into better user experience and high monetization gains.          

\section{Conclusions and Future work}\label{sec:Concluding Remarks and Future work}
This work details the way we incorporated the frequency feature into \offset. Since \offset represents users by their features (a user-less model) it cannot capture individual user CTR drop due to multiple views of the same ads or campaigns by merely training its latent vectors using the logged impressions. Therefore, a rule based HFC was used to prevent users from viewing the same ad too frequently. As an alternative, we considered a SFC approach, where the frequency feature is handled internally by the model and treated as a user-ad feature.  In particular, we set the aggregated ad feature (e.g., campaign), the aggregation time window (e.g., last week), binning vector, and weight category (e.g., global), and trained a frequency weight vector using logistic regression. At serving time we calculate the frequency for the user-ad pairs and add the appropriate weight to the \offset score. Offline and online evaluations demonstrated the significant superiority of the SFC over the legacy HFC ($7.3\%$ CPM lift), which translates into many millions of USD in additional revenue yearly and better user experience. This may be seen as another triumph of machine leaning (ML) over fixed arbitrary man made rules. We note that the SFC enhanced \offset model was pushed into production over 2 years ago and has been serving million of Yahoo users since. 

We also consider at length statistics collected for the frequency feature while the SFC was already deployed, and manage to draw several interesting observations regarding the affect of campaign views frequency over ad click tendency of different crowds. In particular, we show that while both genders are almost equally affected by frequency, the user fatigue (or the decline in CTR with frequency) increases with age.  

The impressive success of the SFC project has sprouted several other ongoing and future projects. Since click tendency with increasing frequency is much different among verticals (see Figure \ref{fig: vertical ctr}) the model by itself might not be able to adjust its predictions accordingly. Therefore, we plan to learn frequency weight vectors per vertical using data collected on that vertical. \Comment{(e.g., learn stream vertical weight vector with data collected on stream vertical only). Another direction considers the category of the weight vectors.} In this work we have found that having a single global vector instead of having one for each ad feature (e.g., campaign) is best due to sparsity issues. These may be eliminated if we consider a hierarchical structure for the weight vectors. We also plan on incorporating recency features (time passed since last view of a campaign) into \offset using a similar soft capping approach.
\balance
\bibliographystyle{ACM-Reference-Format}


\end{document}